\documentstyle[preprint,aps]{revtex}
\begin{document}
\draft

\title{Temperature Dependence of the Spin-Peierls Energy Gap and Anomalous 
Line Shapes in CuGeO$_3$.}

\author{Michael C. Martin$^1$, G. Shirane$^1$, Y. Fujii$^2$, M. Nishi$^2$,
O. Fujita$^3$, J. Akimitsu$^3$, M. Hase$^{4,5}$ and K. Uchinokura$^5$}
\address{$^1$Department of Physics, Brookhaven National Laboratory, Upton, NY
11973-5000\\
$^2$Neutron Scattering Laboratory, Institute for Solid State Physics, 
University of Tokyo, Shirakata, Tokai, Ibaraki 319-11, Japan\\
$^3$Department of Physics, Aoyama Gakuin University, Chitosedai, 
Setagaya-ku, Tokyo 157, Japan\\
$^4$Institute of Physical and Chemical Research (RIKEN), Wako 351-01, Japan\\
$^5$Department of Applied Physics, University of Tokyo, Bunkyo-ku, Tokyo
113, Japan}

\date{12 February 1996, revised 27 March 1996.}
\maketitle

\begin{abstract}
Neutron scattering measurements on a large single crystal of CuGeO$_3$
have been used to determine the temperature-dependence of the spin-Peierls
energy gap.  While the power law behavior of the intensity of structural
superlattice peaks is well fit by $I({\rm T})\propto ({\rm T}_c -
{\rm T})^{2\beta}$ with an exponent of $\beta =0.33$,
the exponent for the temperature dependence of the energy gap 
is significantly smaller than expected for conventional spin-Peierls 
materials.  Usual scaling relations relate the energy gap to the superlattice
reflection intensity as $\Delta({\rm T}) \propto I^a$ with $a=1/3$; the 
present results suggest an exponent of $a\approx 1/6$ for CuGeO$_3$.  
Additional scattering 
cross-section is observed in constant-$q$ and constant-$E$ scans creating a 
long `tail' extending to higher energies relating to a proposed scattering
continuum.  

\end{abstract}

\narrowtext

A spin-Peierls transition in the inorganic compound CuGeO$_3$ was first
reported in 1993 \cite{hase93} and has since been the subject of many 
studies.  This compound has chains of Cu$^{2+}$ ($S=1/2$) along its $c$-axis
and was therefore expected to be fairly one-dimensional in nature.  When
the spin-Peierls transition occurs, the Cu$^{2+}$ chains distort into 
dimers \cite{hirota94} yielding a singlet ground state, and a triplet 
excited state at an energy $\Delta_{SP}$ (the spin-Peierls energy gap) 
\cite{fujita95}.  Neutron 
studies by Nishi {\it et al.} \cite{nishi} demonstrated the existence of 
this gap at about 2 meV at $q=$(0, 1, 0.5) in reciprocal space.  

Evidence is beginning to mount that CuGeO$_3$ is not a typical one-dimensional
spin-Peierls system.  Nishi {\it at al.} \cite{nishi} reported 
spin-wave-like energy dispersions from which the nearest neighbor 
exchange parameters were obtained.  These resulted in a ratio of the
interchain coupling $J'$ to the intrachain coupling $J$ of $J'/J=0.1$.
This is significantly larger than other one-dimensional systems 
($J'/J\sim 1.7\times 10^{-2}$ for CsNiCl$_3$ \cite{CNC} and 
$J'/J \sim 4\times 10^{-4}$ for the Haldane NENP system \cite{NENP}) which
implies that CuGeO$_3$ is not as one-dimensional as was initially thought.
Recent high-pressure neutron scattering measurements \cite{katano,nishipress}
have shown that the dimerizing lattice distortion does not follow the 
spin-Peierls transition temperature and energy gap implying that an 
additional mechanism, such as a spin-only effect, is at play.
Furthermore susceptibility measurements above the 
spin-Peierls transition temperature \cite{hase93} are substantially 
different from the theoretical calculation of Bonner and Fisher 
\cite{bonner} which works well for other one-dimensional $S=1/2$ 
organic spin-Peierls systems.

Since the spin-Peierls ordering is related to a dimerizing lattice distortion,
$\delta$, it is of interest to understand the relationship between this
distortion and the onset of various features of the spin-Peierls phase.
For example the intensity of a superlattice Bragg peak is proportional to
$\delta ^{2}$.  So by fitting the temperature dependent peak intensity data 
to a power-law, one can discover the exponent of the power-law for $\delta$
as a function of temperature.  
Scaling rules have been developed by Cross and Fisher \cite{cross} to relate 
the temperature dependence of the spin-Peierls energy gap ($\Delta_{SP}$) 
to $\delta$ as well.  This relates the energy gap and the superlattice 
intensity such that $\Delta \propto \delta^{2a}$; in Cross-Fisher scaling 
$a=1/3$.  
Alternatively a ground state that is dimerized without a lattice distortion 
was theoretically predicted by Majumdar and Gosh \cite{majumdar} for a 
1D $S=1/2$ antiferromagnet having a nearest neighbor
interaction $J_1$ and a next nearest neighbor interaction $J_2$ related by
$J_1=2J_2$.  Indeed Castilla {\it et al.} \cite{castilla} showed that 
a second neighbor interaction can explain the measured susceptibility and
dispersion curves.  It's possible that the dimerization in CuGeO$_3$ has
origins in both the spin-Peierls spin-phonon mechanism and the latter 
spin-only mechanism.  
We will present the fits we obtain for the energy gap as a function of 
temperature and show that this copper germanate system does not behave as a 
typical spin-Peierls system.  


The primary CuGeO$_3$ single crystal (\#A12) used in the present studies 
was grown by the traveling floating zone method and checked with x-ray 
diffraction and magnetization measurements \cite{nishi}.  It is 
approximately 
$35\times 10\times 4$ mm$^3$ in volume, and has a good mosaic spread in the
$b$ direction, but a poorer mosaic along $c$.  A second, previously studied
sample (\#10) was also used \cite{hirota95}.  It has a smaller volume, 
$27\times 3\times 2$ mm$^3$, and a slightly reduced transition temperature, 
however its mosaic and low background rate are excellent.
The crystals were oriented in an $(h\ k\ h)$ or $(0\ k\ l)$ zone, 
mounted in an aluminum can, and placed in a 3K Displex for
cooling.  Neutron scattering measurements were performed 
at the H7 beam line of the High Flux Beam Reactor at Brookhaven National 
Laboratory in the standard triple-axis configuration.
The spin-Peierls energy gap measurements reported here were obtained
using collimations of 40'-20'-Sample-20'-40', yielding approximately 
a 0.6 meV energy resolution.  The incoming neutrons were monochromated using
the (0 0 2) reflection of pyrolytic graphite (PG) to select 13.7 meV fixed 
incoming energy neutrons, and two PG filters were placed before the sample 
to reduce higher energy contamination.  PG (0 0 2) was also used as an
analyzer.


As reported by Hirota {\it et al.} \cite{hirota94}, structural superlattice
peaks due to the spin-Peierls dimerization are observed at $( h/2, k, h/2)$,
where $h$ is odd and $k$ is any integer.  We plot the intensity of the 
(0.5 3 0.5) reflection as a function of temperature in Figure \ref{IvsT}.
The onset of this peak is well described by a power-law
behavior in the region just below T$_c$.  The solid line in Fig. \ref{IvsT} 
is a fit of the data from 12K - 16K, to a power-law:  $I({\rm T})\propto 
({\rm T}_c - {\rm T})^{2\beta}$.  $\beta$ is the exponent from the 
temperature dependence
of the atomic displacement $\delta$ since $I \propto \delta ^{2}$.  The
fit demonstrates that the 3 dimensional Ising model result of $\beta = 0.33$
is a good description of the intensity data.  This value of $\beta$ is 
somewhat higher than was originally reported by Harris {\it et al.} 
\cite{harris}, 0.24,
which had made Cross-Fisher scaling \cite{cross} appear to hold.
However this report was revised with more accurate measurements to be 0.36
by neutrons \cite{hirota95} and 0.33 by x-rays \cite{harris2}, both consistent
with the present results, which we will show leads to a significant deviation 
from the Cross-Fisher scaling rule.
The T$_c$ is also obtained for this sample from the fit in Figure \ref{IvsT}
to be $14.21\pm 0.02$K, in good agreement with previously
reported values for CuGeO$_3$.  

We studied the spin-Peierls energy gap in the two CuGeO$_3$ crystals.
Figure \ref{Escans} presents the temperature dependent data on the larger 
\#A12 sample, which was aligned in the $(h k h)$ zone, at $q=$ (0.5, 1, 0.5).
While this is not at the zone center, we do not expect a significant
change in the energy gap behavior since the dispersion in the $a$ direction 
is small.  Indeed we will show that the energy gap data
measured on sample \#10 at the zone center, $q=$ (0, 1, 0.5), has the
identical temperature dependence.  The data points of Figure \ref{Escans} 
are the uncorrected data with each successive temperature's data 
set being shifted vertically by 50 counts for clarity.  The
solid lines are fits to two Gaussians, one at zero energy (Bragg) 
and one for the energy gap.  Arrows denote the fitted center positions of
the energy gap at each temperature.  We can clearly observe an energy gap 
peak for temperatures up to 14.0K, after which we can no longer accurately 
fit a peak.  During the course of this experiment we became aware of works
claiming that a ``psuedogap'' feature persists to a few degrees above T$_c$
\cite{lorenzo}.  As seen in Figure \ref{Escans}, our data for temperatures
above 14K do not have any clearly identifiable peak positions.

We summarize the engeries of the gap peaks at all temperatures
measured in Figure \ref{EgapvsT}.  According to the Cross-Fisher scaling 
theory \cite{cross} the onset of the energy gap can be related to the
atomic dimerization displacement such that $\Delta_{SP} \propto \delta^{2a}$
with $a=1/3$.
This theory has been used to successfully fit data of other organic 
spin-Peierls materials \cite{cross} and one would expect that if CuGeO$_3$ is 
a typical spin-Peierls system this Cross-Fisher scaling would again fit the 
observed temperature dependence.  From the data in Fig. \ref{IvsT} we
found that near T$_c$ the exponent $\beta=0.33$ works well.  
Using the Cross-Fisher result, we fit the energy gap data in Figure 
\ref{EgapvsT} to $\Delta_{SP}\propto ({\rm T}_c - {\rm T})^{2\beta a}$ 
with $a=1/3$, 
and obtain the solid line labeled $a=1/3$.  This clearly does not describe the 
energy gap data, even in the vicinity of T$_c$.

By allowing $a$ to be a free fitting parameter, we find that the best fit is 
obtained with $a=0.147 \pm 0.012$.  This value of $a$ is approximately a factor
of two smaller than was expected from Cross-Fisher scaling (1/3).  We therefore
see that to relate the spin-Peierls energy gap in CuGeO$_3$ to the atomic
displacement, the proportionality must be $\Delta_{SP} \propto \delta^{2a}$
with $a\approx 1/6$.

The gap as a function of temperature at the $q=$(0, 1, 0.5)
point in reciprocal space was measured on the second sample (\#10).
This sample's slightly depressed T$_c$ is evident in the top panel of 
Figure \ref{10gap}.  A typical constant-$q$ scan is shown in the inset to 
the lower panel of Fig. \ref{10gap}, and the resultant temperature dependence 
is displayed in the main lower panel.  A power-law fit results in a 
value of $a$ that, within errors, is the same as was found from Fig. 
\ref{EgapvsT}, verifying our above analysis.

As we pointed out in the introduction, the susceptibility of CuGeO$_3$ above
the transition temperature \cite{hase93} was not well fit by the Cross-Fisher 
theory.  A study by Riera {\it et al.} \cite{riera} pointed out
that a much better fit to the susceptibility data could be obtained if
the spin-Peierls energy gap scaled linearly with the atomic dimerization
distortion ($\Delta_{SP} \propto \delta^{2a}$ with $a=1/2$).  However
our result that $a\approx 1/6$ heads in the opposite direction
from that suggested by Riera {\it et al.} \cite{riera}. 
A theoretical understanding of the temperature and pressure 
dependence of the energy gap results, a challenging task, must be pursued.


When we extend the energy gap scans to higher energies, we find that the 
peak has a `tail' with significant scattering intensity at higher energies.
This is shown in Figure \ref{extraintens} and is in agreement
with other recent works \cite{lorenzo,regnault}.  The extra cross-section 
observed at higher energies could be a signature of the theoretically predicted
continuum above the well-defined dispersion \cite{haas}.  
The intensity and width of the energy gap peak have an interesting 
$q$-dependence:  while the integrated intensity (FWHM $\times $ intensity)
is monotonically decreasing with increasing $h$ in support of
the results of a calculation by Haas and Dagotto \cite{haas},
the peak shows a dramatic narrowing near $q=$(0.7, $\bar{1}$, 0.7).
We have made preliminary calculations of the neutron
scattering resolution ellipse projected onto the relevant plane and find that 
its energy-$q$ slope is 14.7 meV\AA \ at (0.5, $\bar{1}$, 0.5) and goes up to 
15.2 meV\AA \ at (0.7, $\bar{1}$, 0.7).  This matches the dispersion near 
$h=0.5$ and $0.75$ whereas the dispersion is considerably steeper in between.
Therefore the narrowing observed near
those wavevectors originates from focusing effects and is not intrinsic
to the sample.

A constant-energy scan, such as the one shown in Figure \ref{continuum}, 
demonstrates that the extra scattering intensity discussed is centered 
around $h=0.5$; instead of two peaks symmetrical around 
$q=$(0.5, $\bar{1}$, 0.5), sizable intensity exists at $h=0.5$.  This
indicates the presence of a continuum of scattering intensity, consistent
with a Fano line shape recently reported from polarized Raman scattering
measurements \cite{vanloos}, and  similar to what has 
recently been observed in KCuF$_3$ \cite{nagler}.  The inset to Fig.
\ref{continuum} shows a sketch of this continuum (in the case of a uniform
chain) and the solid horizontal
line indicates where the constant-energy scan shown in the main portion
of the figure was obtained.
A full analysis and understanding of these observations is yet to be 
completed, and larger and higher quality crystals are required for still 
better counting rates enabling a further exploration of this continuum.
Recently we have been made aware that just such a magnetic continuum is
clearly observed by Arai {\it et al.} \cite{arai} from pulsed neutron
measurements.

{\it Note} After this work was completed we became aware of the results
of Lussier {\it et al.} \cite{lussier} which presents very similar data
to that of our Figure 2.  Although their best fit to a power-law is with an
exponent of $a=0.12$, quite close to our present results, their interpretation
differs somewhat.

\acknowledgments
We would like to thank R.J. Birgeneau, Guillermo Castilla, Vic Emery, Kazu 
Hirota and Kazu Kakurai for informative discussions.  We also thank 
J.E. Lorenzo and collaborators for communicating their results prior to 
publication.  This work was supported in part by the U.S.- Japan Cooperative 
Research Program on 
Neutron Scattering operated by the U.S. Department of Energy and the Ministry
of Education, Science, Sports and Culture (MONBUSHO of Japan), and 
NEDO (New Energy and Industrial Technology Development Organization) 
International Joint Research Grant.
OF and JA were supported by NEDO and by the Grant-in-Aid for Scientific
Research from MONBUSHO of Japan.
Research at Brookhaven National Laboratory was supported by the Division of 
Materials Research at the U.S. Department of Energy, under contract No. 
DE-AC02-76CH00016.


\begin{figure}
\caption{Superlattice peak intensity as a function of temperature showing the
onset of the spin-Peierls state.  The solid line is a power-law fit with the
parameters shown.}
\label{IvsT}
\end{figure}

\begin{figure}
\caption{Constant $q=$(0.5, 1, 0.5) scans at various temperatures near the 
spin-Peierls transition.  The energy gap is clearly
observed up to 14.0K.  Solid lines are fits to Gaussian line shapes.}
\label{Escans}
\end{figure}

\begin{figure}
\caption{The spin-Peierls energy gap, obtained from scans as in Fig. 
\protect\ref{Escans}, plotted as a function of temperature.  The two 
solid lines show the (poor) fit if Cross-Fisher scaling is applicable 
($a=1/3$), and the best fit having an exponent over twice as small.}
\label{EgapvsT}
\end{figure}

\begin{figure}
\caption{$q=$(0.5, 6, 0.5) intensity and $q=$(0, 1, 0.5) energy gap as a 
function
of temperature for sample \#10.  Inset demonstrates a constant-$q$ scan of
the energy gap at T=12.2K.}  
\label{10gap}
\end{figure}

\begin{figure}
\caption{Constant $q=$(0.5, $\bar{1}$, 0.5) scan at T=4.0K showing the 
spin-Peierls energy gap and its non-Gaussian line shape.  The solid line 
is a guide to the eye while the dashed line is a fit to a 
Gaussian line shape.}  
\label{extraintens}
\end{figure}

\begin{figure}
\caption{A constant-energy scan at $E=6.0$ meV around $q=(h,\ \bar{1},\ h)$
at T=4.0K.  The inset sketches a proposed scattering continuum (for a 
uniform spin-${1\over{2}}$ chain) and its 
relation to where the scan was obtained.}
\label{continuum}
\end{figure}

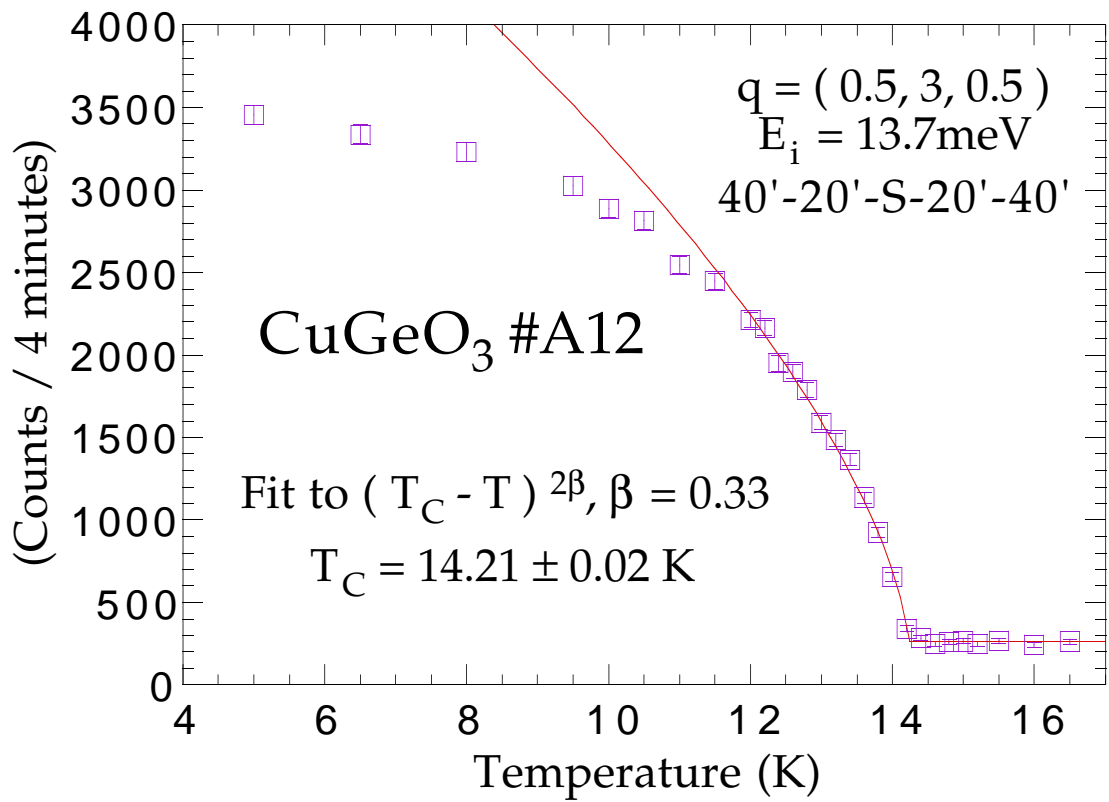
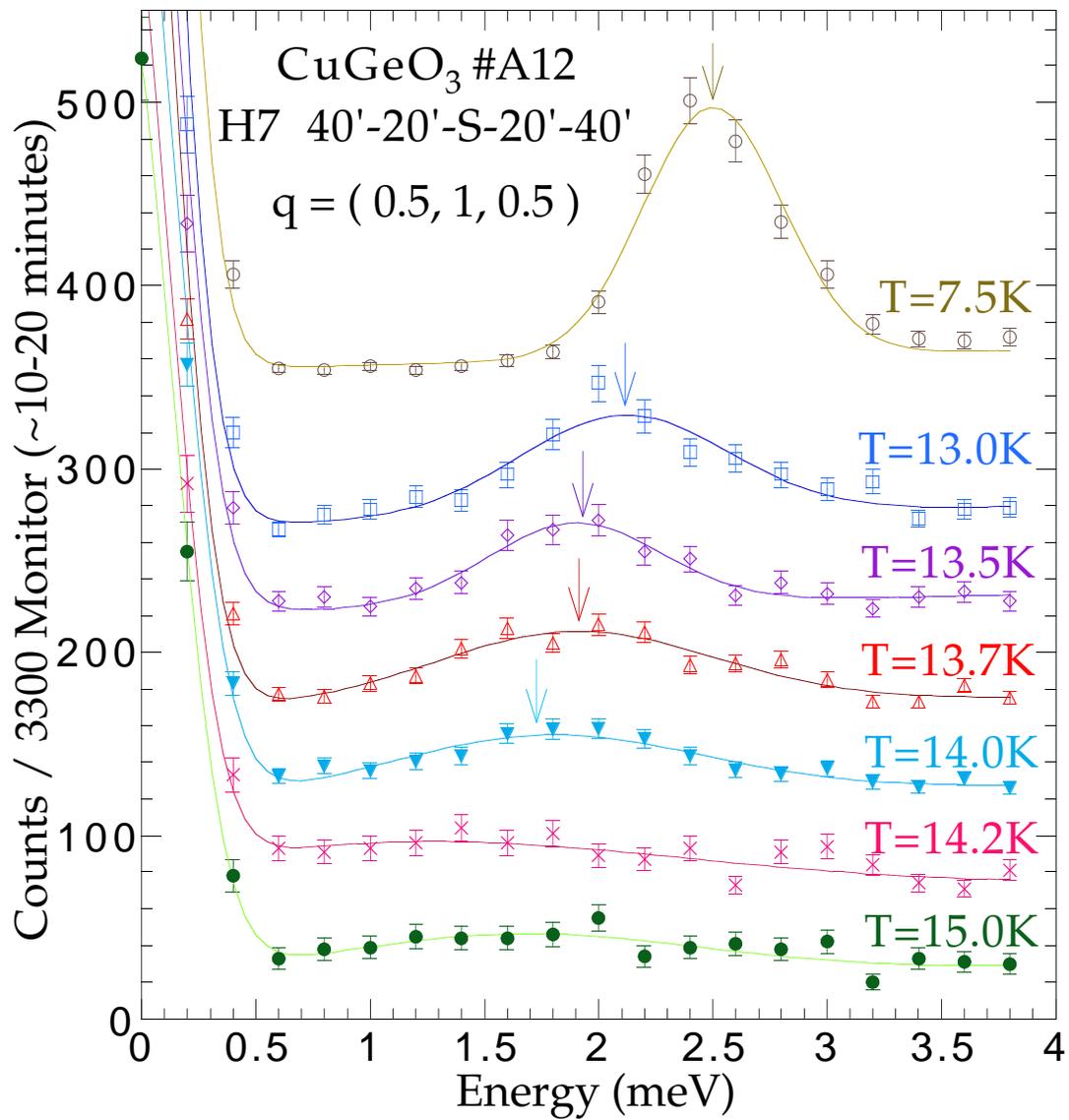
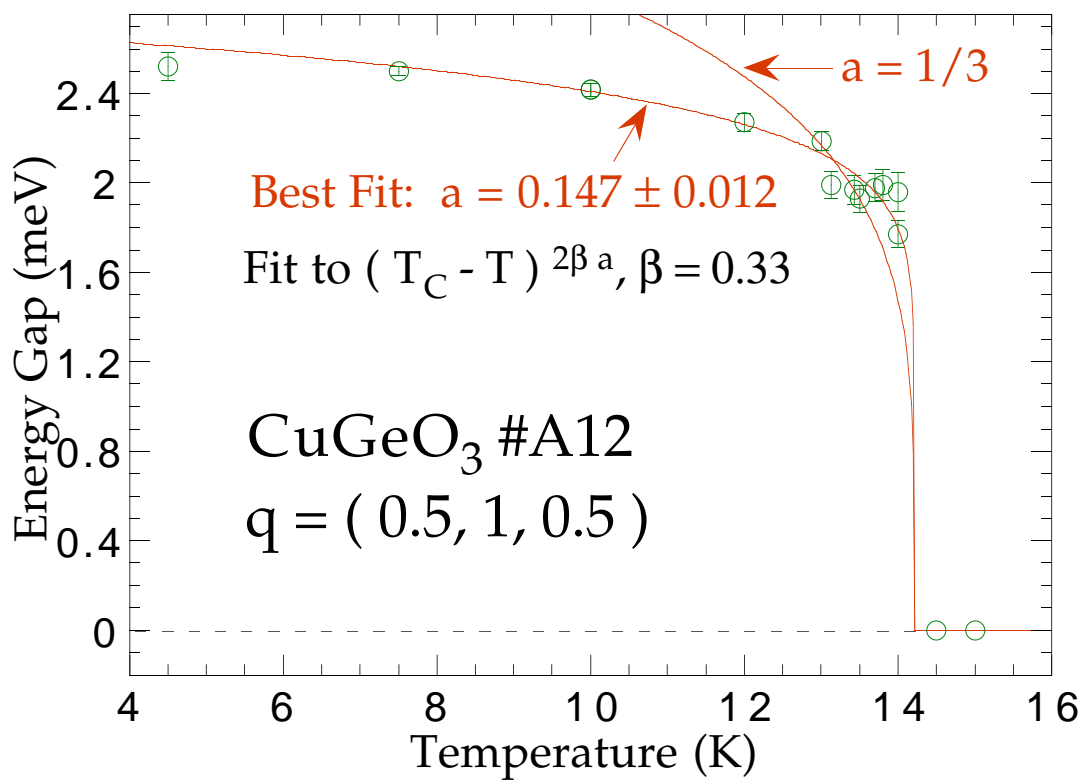
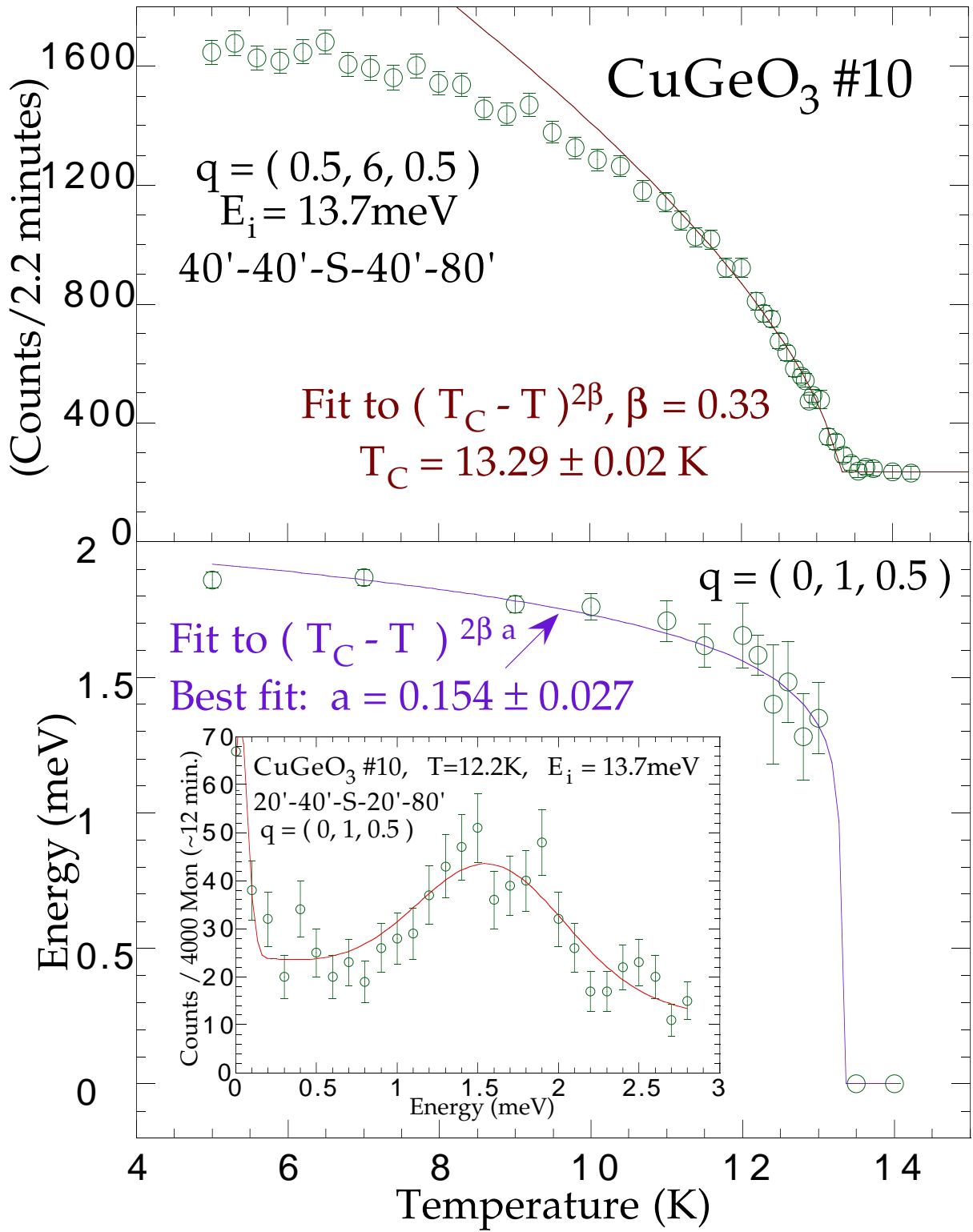
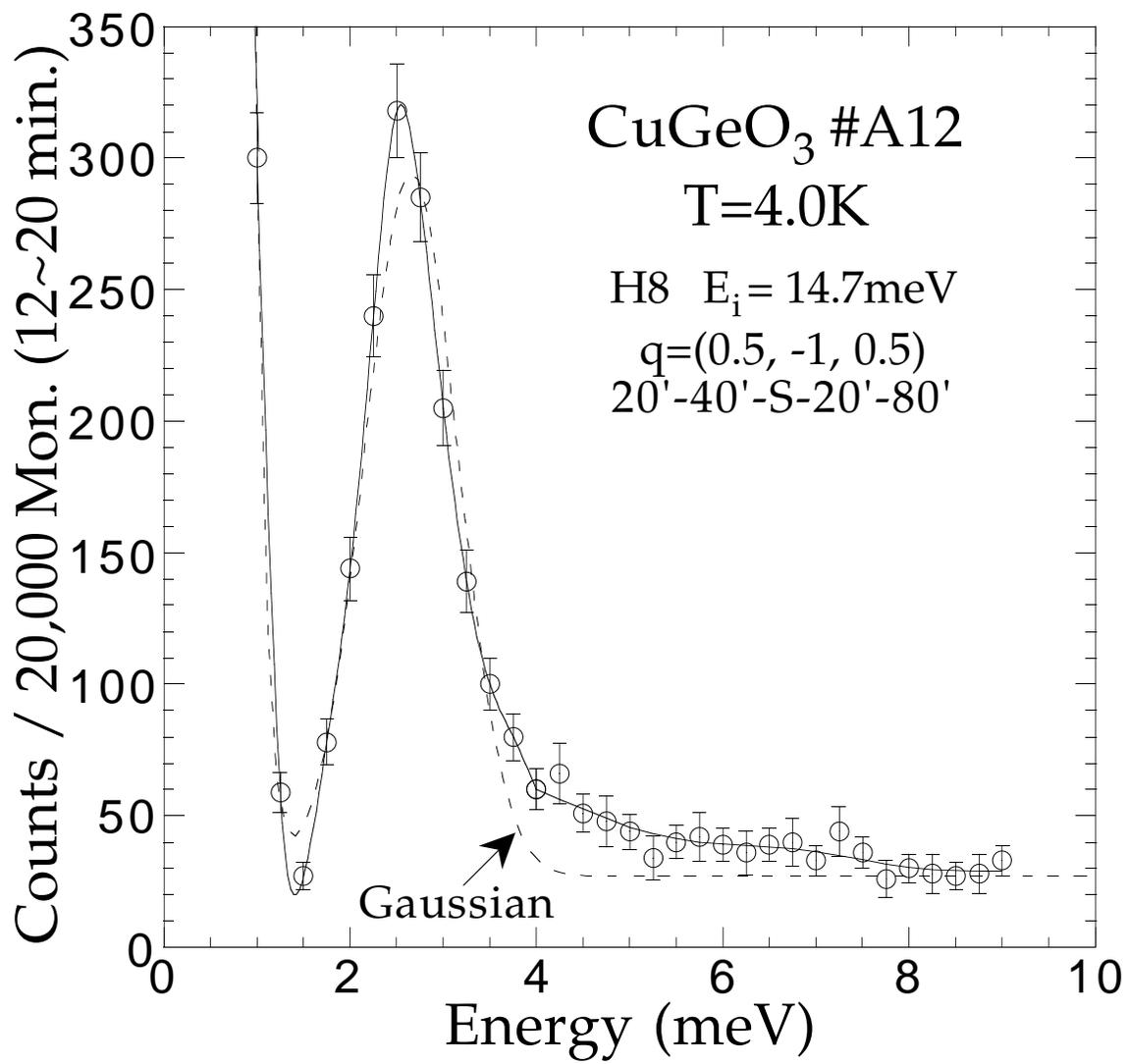
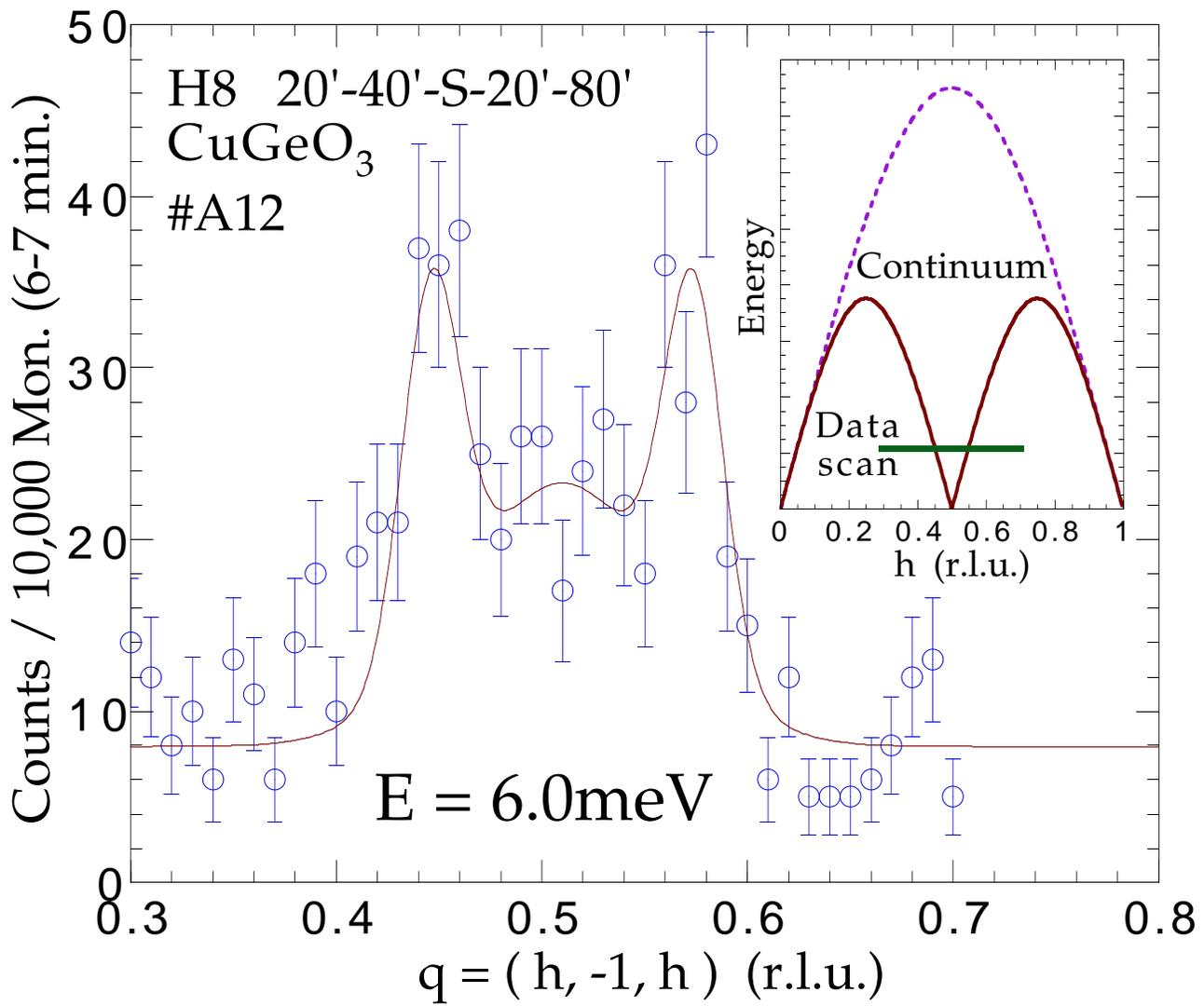
\end{document}